\documentclass[aps,prd,reprint,superscriptaddress,nofootinbib,longbibliography]{revtex4-2}

\usepackage[T1]{fontenc}
\usepackage[utf8]{inputenc}
\usepackage{lmodern}
\usepackage{amsmath,amssymb,bm,mathtools}
\usepackage{graphicx}
\usepackage{booktabs}
\usepackage{hyperref}
\usepackage{xcolor}

\hypersetup{colorlinks=true,citecolor=blue,linkcolor=blue,urlcolor=blue}

\newcommand{\ii}{\mathrm{i}}
\newcommand{\dd}{\mathrm{d}}
\newcommand{\ee}{\mathrm{e}}
\newcommand{\avg}[1]{\left\langle #1 \right\rangle}
\newcommand{\order}[1]{\mathcal{O}\!\left(#1\right)}

\begin{document}

\title{Toward selective quantum advantage in hadronic tomography:\\
explicit cases from Compton form factors, GPDs, TMDs, and GTMDs}


\author{I.~P.~Fernando}
\email{ishara@virginia.edu}
\affiliation{Department of Physics, University of Virginia, Charlottesville, Virginia 22904, USA}

\author{D.~Keller}
\email{dustin@virginia.edu}
\affiliation{Department of Physics, University of Virginia, Charlottesville, Virginia 22904, USA}

\begin{abstract}
We recast the case for quantum advantage in hadronic physics as an observable-by-observable question rather than a blanket claim about Quantum Chromo-Dynamics (QCD). Focusing on hadronic tomography, we analyze why Compton form factors (CFF), generalized parton distributions (GPDs), Transverse Momentum-dependent Distributions (TMDs), and Generalized Transverse Momentum-dependent Distributions (GTMDs) are natural quantum targets: they are defined by light-front, off-forward, or real-time correlation functions whose extraction from Euclidean calculations or sparse experimental data is often an ill-posed inverse problem. We separate three notions of advantage---algorithmic, computational, and representational---and connect each to explicit formal objects. At the algorithmic level, Hamiltonian simulation, linear-response algorithms, and amplitude-estimation primitives motivate gains for real-time and sign-problematic observables. At the computational level, direct quantum evaluation of matrix elements and correlators becomes plausible for PDFs, GPDs, timelike response, and high-energy evolution. At the inference level, recent Quantum Deep Neural Network (QDNN) studies of CFF extraction indicate improved performance in noisy and sparse regimes and motivate hybrid fits in which a quantum simulator supplies a physics prior while a classical network models detector and nuisance effects. We discuss why real-device execution is scientifically necessary, summarize current hardware milestones, and propose benchmark criteria for credible claims of quantum advantage in hadronic tomography.
\end{abstract}

\maketitle

\section{Introduction}

The strongest case for quantum computing in hadronic physics is not that every difficult QCD calculation should migrate to quantum hardware. It is that a specific class of hadronic observables is already quantum-native in its formal definition: light-front bi-locals, off-forward amplitudes, current-current correlators, open-system evolution equations, and real-time scattering amplitudes rather than static Euclidean expectation values. This distinction matters because it separates three questions that are often conflated. One question is whether a quantum algorithm asymptotically outperforms the best known classical strategy. A second question is whether a real device can deliver an end-to-end computational gain after state preparation, compilation, noise, and measurement are included. A third question, especially relevant for data analysis, is whether a quantum or quantum-inspired model gives a representational advantage in a structured inverse problem. All three notions arise in hadronic tomography, but not for the same observables and not on the same timescale \cite{Jordan2012,Jordan2014,Jordan2018,TroyerWiese2005,Funcke2023}.

This article sharpens that claim using explicit examples from Compton form factors, generalized parton distributions, transverse-momentum-dependent distributions, and generalized transverse-momentum-dependent distributions. These quantities sit at the center of the modern hadron-imaging program and also expose the formal reasons that selective quantum advantage is plausible. In the CFF/GPD case, the core task is an underconstrained inversion from finite sets of deeply virtual Compton scattering harmonics to complex amplitudes or nonperturbative functions subject to analyticity, polynomiality, and positivity constraints \cite{Ji1997,Diehl2003,BelitskyRadyushkin2005,GuidalMoutardeVanderhaeghen2013}. In the TMD and GTMD cases, the target object is a Wilson-line correlator that retains transverse momentum, and for GTMDs also off-forward momentum transfer, so the observable is both higher dimensional and more phase sensitive than ordinary PDFs \cite{Bacchetta2007,Meissner2009,LorcePasquini2011}. In the real-time response case, the target is a Minkowskian correlator or structure function whose Euclidean reconstruction is intrinsically delicate \cite{RoggeroCarlson2018,CohenLammLawrenceYamauchi2021,Barata2024}.

The argument of this paper is therefore selective. We do not claim that stable hadron masses are the flagship near-term target; classical lattice QCD already performs extremely well there, and recent work makes the case directly that resonances, timelike observables, finite-density regimes, and other real-time quantities are more promising targets \cite{Funcke2023,Lamm2026}. Instead, we argue that hadronic tomography provides a coherent and technically rich program in which three sources of quantum leverage can be made explicit:
\begin{enumerate}
    \item \emph{algorithmic advantage}, supported by quantum field theory simulation results, {\it Bounded-error Quantum Polynomial time} completeness or BQP-completeness \cite{BernsteinVazirani1993} of scattering-type tasks, and the hardness of the generic fermion sign problem \cite{Jordan2012,Jordan2014,Jordan2018,TroyerWiese2005};
    \item \emph{computational advantage}, when matrix elements or correlators are measured more directly in real time or on the light front than in an Euclidean reconstruction pipeline \cite{Kreshchuk2020,LammLawrenceYamauchi2019,Li2021,GustinGoldstein2022,Barata2024,Agrawal2026};
    \item \emph{representational or inference advantage}, when a quantum feature map, variational circuit, or simulator-defined latent manifold better matches the structure of the inverse problem than a generic classical ansatz \cite{PerezSalinas2021,Schuld2021,LeKeller2025,LeKeller2026}.
\end{enumerate}

The remainder of the paper makes this concrete. Section~\ref{sec:formalism} reviews the formal targets. Section~\ref{sec:advantage} explains why these observables are natural quantum-advantage candidates. Section~\ref{sec:examples} develops explicit examples ranging from CFF extraction to direct light-front correlators and timelike response. Section~\ref{sec:hardware} explains why real devices are scientifically necessary. Section~\ref{sec:benchmarks} proposes practical benchmark criteria.

\section{Formal targets in hadronic tomography}
\label{sec:formalism}

\subsection{CFFs and GPDs in deeply virtual Compton scattering}

At leading twist, quark GPDs are defined by light-front bi-local operators between nonidentical hadron states \cite{Ji1997,Diehl2003,BelitskyRadyushkin2005},

\begin{equation}
\begin{aligned}
&F^q (x,\xi,t) =
\frac{1}{2}\int\frac{\dd z^-}{2\pi}\,
\ee^{\ii x P^+ z^-} \nonumber \\
& \times \left.
\langle p'|\bar q(-z/2)\gamma^+ \mathcal W q(z/2)|p\rangle
\right|_{z^+=0,\bm z_T=0}
\\
&=
\frac{1}{2P^+}\,
\bar u(p')
\left[
\gamma^+ H^q(x,\xi,t)
+\frac{\ii \sigma^{+\mu}\Delta_\mu}{2M}E^q(x,\xi,t)
\right]
u(p),
\end{aligned}
\label{eq:gpd_def}
\end{equation}


with $P=(p+p')/2$, $\Delta=p'-p$, $t=\Delta^2$, and skewness $\xi=-\Delta^+/(2P^+)$. Analogous axial-vector matrix elements define $\widetilde H^q$ and $\widetilde E^q$. GPDs interpolate between ordinary PDFs, elastic form factors, and impact-parameter densities, and they enter Ji's sum rule \cite{Ji1997},
\begin{equation}
J^q = \frac{1}{2}\int_{-1}^{1}\dd x\, x\left[H^q(x,\xi,0)+E^q(x,\xi,0)\right],
\label{eq:ji_sum_rule}
\end{equation}
which ties off-forward tomography directly to quark angular momentum.

In deeply virtual Compton scattering (DVCS) and timelike Compton scattering (TCS), the experimentally fitted quantities are not the GPDs themselves but the complex Compton form factors. At leading order,

\begin{align}
\mathcal H(\xi,t,Q^2)=
\sum_q e_q^2\int_{-1}^{1}\dd x\,
& \left[
\frac{1}{\xi-x-\ii 0}
-\frac{1}{\xi+x-\ii 0}
\right] \nonumber \\
& \times H^q(x,\xi,t;\mu^2),
\label{eq:cff_convolution}
\end{align}

with analogous expressions for $\mathcal E$, $\widetilde{\mathcal H}$, and $\widetilde{\mathcal E}$\cite{BelitskyKirchnerMueller2002,Diehl2003,BelitskyRadyushkin2005}. The imaginary part accesses the diagonal $x=\pm \xi$ combination, while the real part is a principal-value convolution. This already explains why CFF extraction is a difficult inverse problem: experimental observables sample harmonics of the Bethe-Heitler--DVCS interference and DVCS-squared terms, not the GPD directly.

A standard twist-two combination is
\begin{equation}
\mathcal C_{\mathrm{unp}}^{I}
=
F_1 \mathcal H
-\frac{t}{4M^2}F_2 \mathcal E
+\frac{x_B}{2-x_B}(F_1+F_2)\widetilde{\mathcal H},
\label{eq:interference_combo}
\end{equation}
whose imaginary part enters the leading beam-spin harmonic $A_{LU}^{\sin\phi}$ for an unpolarized target \cite{BelitskyKirchnerMueller2002,GuidalMoutardeVanderhaeghen2013}. Even in a restricted local-fit strategy the task is already a sparse, noisy, complex-valued regression problem over correlated kinematics. That makes CFF extraction a natural testbed for quantum-enhanced inference \cite{LeKeller2025,LeKeller2026}.

\subsection{TMDs and multi-scale transverse structure}

The corresponding TMD correlator keeps the transverse momentum of the struck parton explicit,

\begin{widetext}
\begin{align}
\Phi^{[\Gamma]}(x,\bm k_T;\eta)
=
\frac{1}{2}\int\frac{\dd z^-\,\dd^2\bm z_T}{(2\pi)^3}\,
\ee^{\ii xP^+ z^- - \ii \bm k_T\cdot \bm z_T}
\left.
\langle P,S|\bar\psi(0)\Gamma \mathcal W_\eta[0,z]\psi(z)|P,S\rangle
\right|_{z^+=0}.
\label{eq:tmd_def}
\end{align}
\end{widetext}

The gauge link $\mathcal W_\eta$ depends on the process through its staple direction and encodes initial- or final-state interactions, making TMD factorization richer than its collinear counterpart \cite{Bacchetta2007, Arnold2009}. At leading twist one encounters, among others, the unpolarized distribution ($f_1$), helicity distribution ($g_{1L}$), transversity ($h_1$), and the naively time-reversal-odd Sivers ($f_{1T}^{\perp}$) and Boer-Mulders ($h_{1}^{\perp}$) functions. The simplest projection reads
\begin{equation}
\Phi^{[\gamma^+]}(x,\bm k_T)
=
f_1(x,k_T^2)
-\frac{\epsilon_T^{ij}k_T^i S_T^j}{M}
f_{1T}^{\perp}(x,k_T^2)
+\cdots.
\label{eq:tmd_projection}
\end{equation}

For phenomenology, the difficulty is not only the function space itself but also the observable structure. In Semi-Inclusive Deep Inelastic Scattering (SIDIS) \cite{Bacchetta2007} or Drell--Yan (DY) \cite{Arnold2009}, measured structure functions involve convolutions among TMDs, fragmentation functions, hard factors, and soft functions over transverse momenta and multiple scales. Compared with collinear GPD fits, TMD analysis adds an explicit transverse degree of freedom and process-dependent Wilson lines. This makes TMDs plausible candidates for hybrid quantum-assisted fitting strategies, even if direct quantum algorithms for full TMD phenomenology are less mature than those for light-front PDFs or GPD analogues.

\subsection{GTMDs, Wigner distributions, and phase-space structure}

GTMDs are the most differential quark correlators in this hierarchy, retaining both transverse momentum and off-forward momentum transfer,
\begin{equation}
\begin{aligned}
W^{[\Gamma]}_{\lambda'\lambda}(x,\bm k_T,\bm\Delta_T,\xi;\eta)
=
\frac{1}{2}
\int \frac{\dd z^-\,\dd^2\bm z_T}{(2\pi)^3}\,
\ee^{\ii xP^+ z^- - \ii \bm k_T\cdot \bm z_T}
\\
\times
\left.
\langle p',\lambda'|
\bar\psi(-z/2)\Gamma \mathcal W_\eta \psi(z/2)
|p,\lambda\rangle
\right|_{z^+=0}.
\end{aligned}
\label{eq:gtmd_def}
\end{equation}
They are often called ``mother distributions'' because GPDs and TMDs are recovered as limits,

\begin{equation}
\int \mathrm{d}^2 \bm{k}_T\, W \xrightarrow{b_T \xleftrightarrow{FT} \Delta|_{-t}} \mathrm{GPD}
\qquad
\int \dd^2\bm b_T\, W \to \mathrm{TMD}.
\label{eq:gtmd_limits}
\end{equation}

A Fourier transform in transverse momentum transfer yields Wigner or phase-space distributions,
\begin{equation}
\rho^{[\Gamma]}(x,\bm k_T,\bm b_T)
=
\int\frac{\dd^2\bm\Delta_T}{(2\pi)^2}
\,\ee^{-\ii \bm\Delta_T\cdot \bm b_T}
W^{[\Gamma]}(x,\bm k_T,\bm\Delta_T),
\label{eq:wigner_def}
\end{equation}
which combine the position and momentum information contained separately in GPDs and TMDs\cite{Meissner2009,LorcePasquini2011}. In this representation, quark orbital angular momentum can be written as a phase-space moment,
\begin{equation}
L_q^z
=
\int \dd x\,\dd^2\bm k_T\,\dd^2\bm b_T\,
(\bm b_T \times \bm k_T)_z\,
\rho^{[\gamma^+]}(x,\bm k_T,\bm b_T),
\label{eq:oam_wigner}
\end{equation}
making the relation between tomography and orbital dynamics especially transparent\cite{LorcePasquini2011}.

GTMDs are therefore especially relevant to the quantum-advantage discussion. They are high-dimensional, off-diagonal, phase-sensitive quantities whose natural mathematical home is a Hilbert-space or density-matrix description. Even if direct GTMD extraction remains a longer-term goal, the formal structure strongly suggests that overlap estimation, off-diagonal matrix-element measurement, or simulator-defined priors could become more useful here than for simpler static observables.

\begin{table*}[t]
\caption{Why specific tomography observables are plausible quantum targets. ``Status'' distinguishes what is already demonstrated from what remains prospective.}
\label{tab:targets}
\begin{tabular}{p{2.6cm}p{3.3cm}p{3.0cm}p{3.8cm}p{2.7cm}}
\toprule
Observable class & Representative formal object & Main classical bottleneck & Plausible quantum lever & Current status \\
\midrule
CFF extraction from DVCS/TCS & $\mathcal F(\xi,t,Q^2)$ from off-forward GPD convolutions & Sparse/noisy inversion from harmonics to complex amplitudes; correlated uncertainties & QDNN or quantum-kernel regression; simulator-constrained latent manifolds & Hadronic QDNN application exists\cite{LeKeller2025,LeKeller2026} \\
Direct PDF/GPD matrix elements & Light-front bi-locals and hadronic tensors & Real-time or light-cone correlators are awkward in Euclidean methods & Hamiltonian simulation, linear-response algorithms, ancilla overlap estimation & Toy-model algorithms and first hardware PDF\cite{Kreshchuk2020,LammLawrenceYamauchi2019,Li2021,GustinGoldstein2022,Chen2025} \\
TMD phenomenology & $\Phi^{[\Gamma]}(x,\bm k_T;\eta)$ with process-dependent gauge links & High-dimensional multi-scale convolutions and global fits & Hybrid quantum feature maps or simulator priors for transverse structure & Mostly prospective \\
GTMD and Wigner structure & $W^{[\Gamma]}(x,\bm k_T,\bm\Delta_T,\xi)$ and $\rho(x,\bm k_T,\bm b_T)$ & Off-diagonal, phase-sensitive, high-dimensional correlators & Overlap estimation, density-matrix methods, structured priors & Largely prospective\cite{Meissner2009,LorcePasquini2011} \\
Timelike response and small-$x$ evolution & $\avg{J_\mu(t)J_\nu(0)}$; Lindblad evolution of hadronic density matrices & Analytic continuation, sign problems, open-system dynamics & Direct real-time simulation and quantum open-system algorithms & Schwinger-model tests and JIMWLK proposal\cite{Barata2024,Agrawal2026} \\
\bottomrule
\end{tabular}
\end{table*}

\section{Why quantum advantage is plausible for these targets}
\label{sec:advantage}

\subsection{Algorithmic motivation from quantum field theory}

The most rigorous part of the case begins at the level of complexity theory and asymptotic simulation. In Refs. \cite{Jordan2012,Jordan2014} showed that scattering amplitudes and related quantities in quantum field theory can be simulated in time polynomial in the energy, volume, particle number, and inverse target precision, both for scalar and fermionic theories. Related work established the BQP-completeness of a closely related scattering-type task in scalar field theory \cite{Jordan2018}. On the classical side, Refs. \cite{Cook1971,TroyerWiese2005} proved that the generic fermion sign problem is {\it Nondeterministic-Polynomial-time} hard or NP-hard. These results do not imply immediate quantum utility for precision QCD, but they formalize why one should expect selective asymptotic advantage precisely where hadronic observables inherit real-time dynamics, interference, or sign-problem obstructions.

A second algorithmic point is measurement scaling. If the relevant observable can be cast as a probability or amplitude in a coherent quantum computation, Quantum Amplitude Estimation (QAE) offers the familiar improvement
\begin{equation}
N_{\mathrm{queries}}^{\mathrm{QAE}}=\order{\epsilon^{-1}},
\qquad
N_{\mathrm{samples}}^{\mathrm{direct}}=\order{\epsilon^{-2}},
\label{eq:qae_scaling}
\end{equation}
for target additive precision $\epsilon$ in the ideal fault-tolerant setting \cite{Brassard2002}. For hadronic applications, this matters whenever structure functions, correlators, or matrix elements are accessed through coherent overlap-estimation primitives rather than incoherent repeated sampling. It is not a near-term guarantee, but it clarifies what a long-term quantum advantage would look like operationally.

The crucial caveat is that none of these theorems implies blanket advantage for all hadronic observables. Classical lattice QCD already dominates static spectroscopy of stable hadrons, and recent discussions stress that real-time, resonance, finite-density, and entanglement-sensitive observables are much more promising targets than masses \cite{Funcke2023,Lamm2026}. This is precisely why the observable-level analysis of Sec.~\ref{sec:formalism} matters.

\subsection{Observable-level primitives: matrix elements, response functions, and off-diagonal overlaps}

The hadronic quantities introduced above can all be viewed as specializations of a generic task: estimate matrix elements or correlation functions of the form
\begin{equation}
C_{AB}(t)=\langle \psi_0|A^\dagger(t)B(0)|\psi_0\rangle,
\qquad
A(t)=\ee^{\ii H t} A\,\ee^{-\ii H t},
\label{eq:generic_correlator}
\end{equation}
or of off-diagonal form $\langle \psi'|O|\psi\rangle$. Light-front PDFs and GPDs are bi-local correlators with kinematic Fourier transforms; TMDs add transverse separation; GTMDs add momentum transfer between initial and final states; timelike response functions are Fourier transforms of current-current correlators \cite{Ji1997,Diehl2003,Bacchetta2007,Meissner2009,RoggeroCarlson2018,Barata2024}. A quantum device is attractive because it can, in principle, prepare $|\psi_0\rangle$, evolve it unitarily in real time, and estimate the desired overlap interferometrically. By contrast, the Euclidean strategy often computes an imaginary-time correlator and then solves an unstable inverse problem to recover a Minkowskian object.

This is the sense in which hadronic tomography is more naturally aligned with quantum simulation than stable masses are. The key issue is not simply state-space dimension; it is the mismatch between the native definition of the observable and the most efficient classical route. Light-front and timelike quantities are not impossible classically, but they are often indirect. Quantum hardware offers a route that is closer to the definition itself.

\subsection{Inference-level advantage and simulator-defined priors}

The same logic appears at the data-analysis level. A CFF fit does not merely interpolate smooth functions. It reconstructs a structured latent object constrained by analyticity, symmetry, kinematic support, and often by cross-channel relations. If a quantum feature map or variational circuit captures that latent geometry more economically than a generic classical network, then the relevant advantage is representational rather than asymptotic. This is exactly the perspective advocated by the modern quantum machine-learning literature: the benefit, when it appears, comes from the induced feature space or kernel, not from calling every parametrized circuit a neural network in the classical sense \cite{PerezSalinas2019,Schuld2021}.

This point becomes sharper when the quantum component is treated as a physics prior rather than as a standalone regressor. A useful generic architecture is
\begin{equation}
\mathcal M_i(\theta,\phi)
=
\mathcal N_i\!\left(x_i;\phi,\mathcal O_i^{(q)}(\theta)\right),
\label{eq:model_architecture}
\end{equation}
where $\mathcal O_i^{(q)}(\theta)$ denotes a matrix element, correlator, or amplitude generated by a gauge-aware quantum circuit or simulator, while $\mathcal N_i$ is a classical nuisance model handling detector response, acceptance, binning, or residual model discrepancy. One then fits
\begin{align}
\chi^2(\theta,\phi)=
\sum_{ij}
\Big[y_i^{\mathrm{exp}}-\mathcal M_i(\theta,\phi)\Big]
&\Sigma_{ij}^{-1}
\Big[y_j^{\mathrm{exp}}-\mathcal M_j(\theta,\phi)\Big] \nonumber \\
& +\lambda\,\mathcal R(\theta,\phi),
\label{eq:hybrid_chi2}
\end{align}
with covariance matrix $\Sigma$ and regularizer $\mathcal R$. Equations~\eqref{eq:model_architecture} and \eqref{eq:hybrid_chi2} are not tied to a specific implementation; they formalize the idea that the quantum subroutine should supply the part of the model where the Hilbert-space structure matters most, while classical layers absorb nuisance structure. For hadronic tomography, this is a more realistic near-term path than an end-to-end claim that a hardware QNN alone will outperform the best classical fit.

\section{Explicit examples}
\label{sec:examples}

\subsection{CFF extraction as a structured inverse problem}

The recent CFF extraction study \cite{LeKeller2025}  provides a direct hadronic example of inference-level quantum advantage. Using pseudodata based on Jefferson Lab DVCS measurements and the twist-two Belitsky--Kirchner--M\"uller formalism, they compared classical deep neural networks with quantum-inspired deep neural networks in a local-fit-like extraction strategy. Their conclusion was not that quantum models dominate universally, but that the QDNN gave improved predictive accuracy and tighter extraction performance at comparable model complexity. The associated follow-up study introduced a quantitative ``quantum qualifier'' and showed that regions favoring the quantum model expand with data complexity, noise, and dimensionality \cite{LeKeller2026}.

This example is important because it mirrors the structure of the formal problem. The fit target is a complex CFF combination like Eq.~\eqref{eq:interference_combo}, inferred from a finite set of harmonics and cross sections with correlated uncertainties. The data are sparse in $x_B$, $Q^2$, and $t$, while the latent object is constrained by the GPD formalism. Such problems are exactly where one should expect a feature-space or latent-manifold advantage to matter if it exists. The quantum qualifier result therefore supports a selective claim: quantum-inspired or hybrid models become more compelling as the inverse problem becomes noisier and more structured, not because quantum circuits are magical but because the induced representation can better match the target geometry.

There is also an instructive connection to earlier work on proton PDFs. Variational quantum circuits have been proposed \cite{PerezSalinas2021} for PDF representation, and explored both simulation-based and real-device deployments in collider PDF fits. Although that work focused on collinear PDFs rather than CFFs, it already framed partonic structure extraction as a variational quantum-learning problem. The CFF application is a more demanding descendant of the same idea because the target amplitudes are complex and off-forward.

\subsection{Direct light-front matrix elements and partonic observables}

The computational case becomes stronger when one asks the quantum device to compute the matrix element itself rather than merely to fit its reconstruction. For ordinary PDFs, the defining light-front correlator is
\begin{equation}
q(x)=
\int \frac{\dd\lambda}{4\pi}\,
\ee^{\ii \lambda x}
\langle P|
\bar\psi(0)\gamma^+ \mathcal W(0,\lambda n)\psi(\lambda n)
|P\rangle,
\label{eq:pdf_def}
\end{equation}
and GPDs generalize this expression off-forward while TMDs and GTMDs add transverse separation and momentum transfer. The common feature is that these are light-cone or real-time correlators, which is exactly why they are difficult to compute directly in Euclidean lattice QCD.

Several quantum approaches now attack this obstacle more natively. Light-front Hamiltonian quantum simulation has been proposed \cite{Kreshchuk2020} for partonic observables. Ref. \cite{LammLawrenceYamauchi2019} argued that PDFs and hadronic tensors are natural quantum-computing targets and emphasized that fitting the hadronic tensor may be the cleanest route to the PDF itself. A hybrid quantum algorithm has been developed \cite{Li2021} for deep-inelastic structure functions and partonic collinear structure. Gustin and Goldstein then extended the light-front program to the computation of GPD analogues in quantum field theory, highlighting the favorable qubit scaling of the light-front formulation \cite{GustinGoldstein2022}. In parallel, tensor-network calculations have demonstrated direct extraction of PDFs and distribution amplitudes from hadronic matrix-product states, providing an essential classical benchmark for any quantum-circuit implementation \cite{KangMoranNguyenQian2025}.

The most concrete milestone to date is the first PDF extraction on a quantum computer, as reported in \cite{Chen2025}. In a Schwinger-model calculation on IBM hardware, they used ten-qubits and one ancilla to reconstruct a light-cone PDF from direct quantum measurements. The current errors are still large, but the central values agree well with classical simulation and the experiment establishes the full operational pipeline: state preparation, real-device evolution, ancilla-based measurement of the relevant correlator, and reconstruction of the distribution function. That is already qualitatively different from a purely classical emulator.

From the perspective of this article, the key point is that Eqs.~\eqref{eq:gpd_def}, \eqref{eq:tmd_def}, \eqref{eq:gtmd_def}, and \eqref{eq:pdf_def} share a common structure. They are all bi-local matrix elements with kinematic Fourier transforms. Once a quantum workflow exists for one member of this family, extensions to GPDs, TMDs, or GTMDs become a matter of additional momentum transfer, Wilson-line structure, and state-preparation overhead rather than a completely new concept. That is why direct GPD or GTMD measurements remain prospective but are not merely speculative.

\subsection{Timelike response, transport, scattering, and small-x evolution}

A third class of examples arises when the target observable is explicitly real-time. Linear response on a quantum computer has been formulated in general terms \cite{RoggeroCarlson2018}, and Ref. \cite{CohenLammLawrenceYamauchi2021} showed how transport coefficients in gauge theories can be obtained without simulating an entire heavy-ion collision, thereby isolating a class of phenomenologically important observables with smaller spacetime volume requirements. These constructions matter for hadronic physics because they shift the goal from complete event simulation to experimentally relevant response functions.

The timelike Hadronic Vacuum Polarization (HVP)  and Hadronic Light-by-Light (HLbL) program is a particularly clear case. Ref. \cite{Barata2024} emphasized that direct timelike first-principles calculations remain difficult in conventional lattice workflows, and used Schwinger-model tests with tensor networks and digital quantum emulation to outline a route toward timelike HVP/HLbL on quantum devices. Here the appeal is not only speed. It is that the target quantity is Minkowskian from the outset.

Real-time hadron dynamics and scattering provide the strongest current hardware illustrations. Farrell \emph{et al.} prepared and propagated hadron wave packets in the Schwinger model using 112 qubits on IBM's Heron architecture \cite{Farrell2024}. Refs. \cite{Davoudi2024,Davoudi2025} demonstrated efficient preparation of interacting mesonic wave packets on trapped-ion hardware and then used related constructions for digital quantum computation of hadron scattering in a lattice gauge theory. The first observation of scattering in a lattice gauge theory on IBM hardware has been reported in \cite{Schuhmacher2025}, including tunable post-collision dynamics in $[1+1]$-D QED. These are still simplified theories, but they show that state-sensitive scattering dynamics can be executed on real devices.

Finally, the high-energy/small-$x$ frontier has acquired a direct quantum formulation. A quantum algorithm for JIMWLK evolution has been proposed \cite{Agrawal2026} by exploiting its reformulation as a Lindblad evolution equation for the hadronic density matrix. Conceptually this is important because it extends the quantum-advantage discussion beyond pure-state wave functions to open-system evolution of a hadronic density operator. That is exactly the direction one expects to matter for saturation physics and future Electron-Ion Collider phenomenology.

\section{Why running on a real quantum device matters}
\label{sec:hardware}

If the goal were only to test mathematical ideas, classical simulators would often suffice. But hadronic applications are unusually sensitive to operational details that only become meaningful on hardware: gauge-symmetry preservation under native compilation, leakage out of the physical subspace, measurement cost under limited budgets, and the extent to which state preparation dominates the total workflow. A classical simulator can validate a circuit identity; it cannot validate whether the quantum protocol remains scientifically useful once those effects are included.

This is particularly acute for tomography observables. Off-forward matrix elements require state preparation for multiple momentum sectors and controlled overlap estimation. Timelike and transport observables require sufficiently coherent real-time evolution. TMD and GTMD extensions would additionally require faithful handling of transverse degrees of freedom and Wilson-line structure. A serious claim of computational advantage must therefore include the costs of state preparation, mitigation, and verification, not just the asymptotic cost of the idealized unitary.

The current hardware literature already illustrates why this matters. Wave-packet preparation on Quantinuum and IBM hardware shows how symmetry-aware ans\"atze and mitigation protocols affect fidelity \cite{Davoudi2024,Farrell2024}. The first hardware PDF computation shows that reducing the two-qubit gate depth to the few-hundred level is a practical threshold for obtaining a sensible light-cone correlator on current devices \cite{Chen2025}. Scattering experiments reveal that early-time observables can be robust while late-time dynamics are limited by decoherence \cite{Davoudi2025,Schuhmacher2025}. Finite-density simulations on trapped ions show that gauge-invariant measurements and thermal-state preparation can be executed together on real devices \cite{Than2025}. These are not marketing milestones; they are the data needed to understand which hadronic workflows scale realistically.

\begin{table*}[t]
\caption{Representative real-device milestones relevant to the hadronic-tomography program.}
\label{tab:hardware}
\begin{tabular}{p{1.0cm}p{4.0cm}p{2.5cm}p{7.6cm}}
\toprule
Year & Demonstration & Platform & Why it matters \\
\midrule
2024 & Interacting mesonic wave-packet preparation in confining gauge theories\cite{Davoudi2024} & Quantinuum H1-1 & Establishes efficient preparation of scattering initial states directly in the interacting theory, avoiding long adiabatic ramps. \\
2024 & Hadron dynamics in the Schwinger model using 112 qubits\cite{Farrell2024} & IBM Heron (\texttt{ibm\_torino}) & Demonstrates localized hadron wave packets above a confining vacuum and real-time propagation with a nontrivial two-qubit-gate budget. \\
2025 & First quantum computation of a PDF on hardware\cite{Chen2025} & IBM quantum computers & Executes the complete light-cone correlator workflow for a PDF, including ancilla-based measurement and reconstruction. \\
2025 & Two-hadron scattering in a lattice gauge theory\cite{Davoudi2025} & IonQ Forte & Shows precision dependence on high-fidelity wave-packet preparation and demonstrates early-time collision dynamics. \\
2025 & First observation of scattering in a lattice gauge theory on hardware\cite{Schuhmacher2025} & IBM \texttt{ibm\_marrakesh} & Demonstrates real-time post-collision dynamics with tunable topological and mass effects in a gauge theory. \\
2025 & Finite-density SU(2) and SU(3) gauge-theory thermal states\cite{Than2025} & Trapped-ion quantum computer & Targets the regime historically associated with the sign problem and validates gauge-invariant measurements in finite-density studies. \\
\bottomrule
\end{tabular}
\end{table*}

A further reason to use real devices is architectural learning. It is already clear from the field-theory literature that different encodings, gauge-fixing strategies, truncations, and symmetry-preserving constructions lead to very different circuit depths and error sensitivities \cite{Funcke2023}. Hadronic tomography will need that kind of architectural experimentation even more than static spectroscopy, because the relevant observables are not protected by the forgiving nature of imaginary-time projection.

\section{Benchmark criteria for a credible claim of quantum advantage}
\label{sec:benchmarks}

A useful hadronic benchmark should not ask whether a quantum routine beats a straw-man classical baseline. It should ask whether the quantum route is superior for a fixed observable, fixed uncertainty target, and fixed total workflow cost. In practice, that means at least the following.

First, the observable should be genuinely native to a quantum workflow: real time, light front, timelike, finite density, open-system, or entanglement dominated. That requirement is what differentiates tomography and scattering from stable mass calculations \cite{Funcke2023,Lamm2026}. Second, the encoding should be symmetry-aware. Gauge invariance, charge sectors, and support constraints should be built into the representation whenever possible rather than enforced as weak penalties after the fact. Third, the classical baseline must be serious: tensor networks, exact diagonalization where available, Euclidean lattice or Monte Carlo methods where applicable, and tuned classical ML baselines for inverse problems \cite{KangMoranNguyenQian2025,Schuld2021,LeKeller2025}. Fourth, total resource accounting must be end to end, including state preparation, compilation, readout, error mitigation, and uncertainty quantification, not just idealized gate counts. Finally, the observable should deliver a physics-level payoff, such as improved posterior constraints, reduced uncertainty on a real amplitude, or direct access to a correlator that was previously indirect.

These criteria are especially important for hybrid workflows. A hybrid fit may not offer asymptotic advantage in the strict complexity-theory sense, but it can still offer computational or representational advantage if the quantum component materially improves posterior coverage, robustness to sparsity, or extrapolation fidelity at fixed total cost. For hadronic tomography this is a feature, not a weakness. The most realistic near-term strategy is not ``replace the entire pipeline by a quantum circuit.'' It is to let the quantum component handle the part of the problem where Hilbert-space structure matters most and let classical machinery absorb the nuisance structure.

\section{Outlook}

The formalism itself suggests where the field should go next. In CFF extraction, the immediate path is to move from quantum-inspired QDNN benchmarking to hybrid fits in which the latent space is explicitly constrained by analyticity, polynomiality, and simple GPD priors. In direct matrix-element computation, the next step is to generalize hardware PDF workflows to more differential observables, especially off-forward and transverse-resolved correlators. In real-time response, the task is to connect linear-response and transport algorithms more directly to observables of hadronic interest, including timelike response and small-$x$ evolution. Across all of these, real hardware must remain part of the loop because state preparation, symmetry protection, and measurement cost are the practical variables that will determine whether selective quantum advantage survives contact with experiment.

The field should also resist two symmetric errors. One is overclaiming: the existence of a quantum algorithm or a toy-model speedup does not imply utility for full QCD on current hardware. The other is underclaiming: because stable hadron masses are not the best target, it does not follow that hadronic physics lacks quantum-advantage opportunities. On the contrary, hadron tomography may be one of the cleanest places in high-energy physics where the observable, the computation, and the architecture can all be aligned.

\section{Conclusions}

The case for quantum advantage in hadronic physics is strongest when framed selectively and at the level of explicit observables. Compton form factors, generalized parton distributions, TMDs, GTMDs, timelike response functions, and small-$x$ density-matrix evolution are natural quantum targets because they are defined by light-front, off-forward, or real-time correlators that are either awkward to reconstruct in Euclidean methods or difficult to infer from sparse and noisy data.

For these targets, three notions of advantage can be formalized. Algorithmic advantage is motivated by known complexity-theory and quantum-field-theory simulation results. Computational advantage is plausible when the quantum device measures matrix elements or correlators closer to their native definition than classical workflows do. Representational advantage arises when quantum or hybrid models provide a better latent geometry for a structured inverse problem, as recent CFF extraction studies suggest.

The main practical conclusion is therefore not that quantum methods should replace all hadronic computation. It is that hadronic tomography offers a disciplined and testable program for \emph{selective} quantum advantage. The most credible near-term route is hybrid: use symmetry-aware quantum circuits or simulators to generate states, correlators, or structured priors, and use classical layers to absorb nuisance structure and perform uncertainty-aware inference. When framed this way, the argument for running on real quantum devices is not promotional but scientific: only hardware can tell us whether the relevant workflows remain useful once compilation, noise, and measurement are taken seriously.

\bibliographystyle{apsrev4-2}
\bibliography{ref}

\end{document}